\documentclass{aa}%[referee] voor de referee
\usepackage{psfig}
\usepackage{psfrag}
\usepackage{latexsym}
\usepackage{graphicx}%[draft] voor {graphicx} laat plaatjes weg
\usepackage{times,amssymb,lscape,verbatim}
\usepackage{natbib}
\bibpunct{(}{)}{;}{a}{}{,} %voor A&A stijl

\begin{document}
\title{RXTE observations of the dipping low-mass X-ray binary 4U~1624--49}
%\subtitle{}

\author{Dave Lommen
	\and Steve van Straaten
	\and Michiel van der Klis
	\and Brechtje Anthonisse
	}

\institute{Astronomical Institute ``Anton Pannekoek'', University of Amsterdam, and Center for High Energy Astrophysics, Kruislaan 403, 1098 SJ Amsterdam, The Netherlands}

\offprints{Dave Lommen,\\ \email{djplomme@science.uva.nl}}

\date{Received ?? / Accepted ??}

% Received 8 March 2004
% Accepted 15 February 2005
% Astro-ph version 26 February 2005
% 4 April 2005: typos fixed

\abstract{
We analyse $\sim 360$ ks of archival data from the \emph{Rossi X-Ray Timing 
Explorer} (\emph{RXTE}) of the 21 hr orbital period dipping low-mass X-ray binary 4U~1624--49. We find that outside the dips the tracks in the colour-colour and hardness-intensity diagrams (CDs and HIDs) are
reminiscent of those of atoll sources in the middle and upper parts of the banana branch.
The tracks show secular shifts up
to $\sim 10\%$.
We study the power spectrum of 4U~1624--49 as a
function of the position in the CD. This is the first time power spectra of this source are
presented. No quasi-periodic oscillations (QPOs) are found. 
The power spectra are
dominated by very low frequency noise (VLFN), characteristic for atoll
sources in the banana state, and band limited noise (BLN) which is not reliably detected but may, uncharacteristically, strengthen and increase in frequency with spectral hardness. The VLFN fits to a power law, which becomes steeper when the source
moves to the harder part of the CD.%, confirming that 4U~1624--49 is an atoll source, which in our observations is always in the banana branch but never reaches the lower left
We conclude that 4U~1624--49 is an atoll source which in our observations is in the upper banana branch.
%banana. 
Combining this with the high (0.5--0.7 $L_{\rm Edd}$) luminosity, the long-term flux stability of the source as seen with the \emph{RXTE All-Sky Monitor} (\emph{ASM}), and with the fact that it is an X-ray dip source, we conclude that 4U~1624--49 is most likely a GX atoll source such as GX 3+1 and GX 9+9, but seen edge on. 
}

\authorrunning{Dave Lommen et al.}
\titlerunning{RXTE observations of the LMXB 4U~1624--49}
\maketitle

\section{Introduction}
\label{sec:introduction}
The low-mass X-ray binary 4U~1624--49 was dubbed The Big Dipper because of the presence
of periodic, 6 to 8 hr duration intensity dips in the X-ray lightcurve \citep{watson:1985}. No periodic dips
longer than $\sim 1$ hr had been observed in other sources. This dipping is thought to be due to occultations of
the central source by a thickened region of the accretion disc rim where the gas stream
from the companion impacts upon the outer disc \citep[see e.g.][]{white:1982}. \citet{frank:1987} suggested that the dips 
are
due to cold clouds closer to the centre of the disc. However, their model predicts little dip activity for photon
energies $\gtrsim$ 6 keV, whereas the dips in 4U~1624--49 are detected up to 15 keV and even discernible above 15 keV
\citep{smale:2001}. The
orbital period of 4U~1624--49 is exceptionally long, 21 hr or 5--25 times longer
than that of other dipping sources. This corresponds to a much larger stellar separation and
accretion disc radius, and it can be assumed to also account for the long duration of the dips.

4U~1624--49 also shows strong flaring activity on timescales of a few thousand seconds. While
dipping is most clear in the energies below $\sim 10$ keV, flaring is not significant at energies below 8 keV \citep{smale:2001}.
\citet{balucinska:2001} presented a hardness-intensity diagram exhibiting a branch which they
interpreted as similar to a Z-source flaring branch \citep[see][for definitions]{hasinger:1989}.

In atoll sources in the banana
state, the very low
frequency noise (VLFN) can be fitted with a power-law shape, $P(\nu) \propto \nu^{-\alpha}$ \citep[see][]{hasinger:1989}, sometimes combined with a
Lorentzian \citep[e.g.][]{reerink:2005,disalvo:2003}. The steepness of the power law increases as the source moves
from the lower to the upper part of the banana. In this paper we study the colour-colour and hardness-intensity diagrams (CDs and HIDs) of 4U~1624--49 and look for 
features in the power spectra as a function of location in the CD.  By investigating in particular the VLFN component, we find that 4U~1624--49 is consistent with being an atoll source which is in the banana state at the time of the
observations.

\section{Observations}
\label{sec:Observations}
In this work we analyse observations from the public \emph{RXTE} archive. The
data are from four proposals, spanning a period from January 1997 to November
1999. The log of these observations is
presented in Table~\ref{tab:observaties}. We used data from the Proportional Counter Array
(PCA; see \citet{zhang:1993} for instrument information, \citet{jahoda:1996} for in-orbit performance) on board 
\emph{RXTE}, which consists of five co-aligned Proportional
Counter Units (PCUs), sensitive in the energy range $2-60$ keV, with a total effective
area of approximately 6,500 cm$^2$ and a field of view, delimited by collimators, of $1^\circ$ FWHM.
Each PCU contains three detector layers and in our analysis we used all the photons collected by the three layers taken together.
We excluded data for which the source was less than 10 degrees above the horizon from
the point of view of the satellite, and data
for which the pointing offset was greater than 0.02 degrees. 
%We used Good Xenon data to create lightcurves with a resolution of 1 s, in which we found bins which contained no data at all (``drop outs''). Assuming this to be 
%corrupt data, we excluded these bins and the ones directly preceding and following 
%them.
We also removed data drop outs and 1 s of data directly preceding and following them.
\begin{table*}
\caption[]{\emph{RXTE} observations of 4U~1624--49 used in this paper. The observations from
P20068 and P20067 are from January 5 and 6 1997, respectively, and are taken together 
in our analysis. The observations from P20071 span 8 days in May 1997, and the 
observations from P30064 span 4 days in 1999. The count rate is taken in
the energy range 2-60 keV. Typical background is 100 counts per second (c/s). The table also shows whether a certain observation contained \emph{no} dip, \emph{part} of a dip, or a \emph{complete} dip. Observation 30064-01-03-00 is not used in our analysis, since it consists mainly of a dip.}
\footnotesize
\label{tab:observaties}
\begin{tabular}{lcrrcc}
\hline \hline
Observation ID & Begin Time     & Good                & Rate$^{\mathrm{b}}$ & Gain  & Dip?     \\
	       & (UTC)	        & Time$^{\mathrm{a}}$ & (c/s)               & Epoch &          \\
	       &                & (sec)               &                     &       &          \\
\hline
20068-01-03-00 & 05/01/97 00:22 & 2820                & 926                 & 3     & no       \\
20068-01-04-00 & 05/01/97 03:44 & 1620                & 997                 & 3     & no       \\
20068-01-05-00 & 05/01/97 07:36 & 1080                & 1039                & 3     & no       \\
20068-01-06-00 & 05/01/97 10:52 & 914                 & 994                 & 3     & no       \\ 
20068-01-07-00 & 05/01/97 14:19 & 2700                & 971                 & 3     & no       \\
20067-01-01-000& 06/01/97 11:14 & 25912               & 1021                & 3     & no       \\
20067-01-01-00 & 06/01/97 18:26 & 5468                & 1092                & 3     & no       \\
20067-01-01-01 & 06/01/97 20:44 & 24250               & 960                 & 3     & no       \\
20071-02-01-00 & 04/05/97 15:57 & 19477               & 943                 & 3     & no       \\
20071-02-01-02 & 06/05/97 10:46 & 5334                & 1001                & 3     & no       \\
20071-02-01-01 & 06/05/97 17:49 & 11720               & 928                 & 3     & no       \\
20071-02-01-03 & 11/05/97 18:47 & 16828               & 890                 & 3     & no       \\
30064-01-02-03 & 27/09/99 18:22 & 1269                & 865                 & 4     & no       \\
30064-01-01-09 & 27/09/99 19:58 & 20529               & 990                 & 4     & no       \\
30064-01-01-04 & 28/09/99 03:09 & 14792               & 1074                & 4     & no       \\
30064-01-01-08 & 28/09/99 09:13 & 15376               & 912                 & 4     & part     \\
30064-01-01-000& 28/09/99 14:09 & 28800               & 975                 & 4     & part     \\
30064-01-01-00 & 28/09/99 22:09 & 2465                & 1108                & 4     & no       \\
30064-01-01-02 & 28/09/99 23:18 & 22084               & 1067                & 4     & no       \\
30064-01-01-03 & 29/09/99 07:32 & 13618               & 836                 & 4     & part     \\
30064-01-01-010& 29/09/99 12:30 & 28800               & 987                 & 4     & no       \\
30064-01-01-01 & 29/09/99 20:30 & 12424               & 1025                & 4     & no       \\
30064-01-01-050& 30/09/99 00:45 & 23997               & 916                 & 4     & part     \\
30064-01-01-05 & 30/09/99 07:25 & 15547               & 974                 & 4     & no       \\
30064-01-01-06 & 30/09/99 12:51 & 20077               & 1059                & 4     & no       \\
30064-01-01-07 & 30/09/99 23:22 & 2300                & 857                 & 4     & no       \\
30064-01-02-01 & 01/10/99 00:01 & 1200                & 881                 & 4     & no       \\
30064-01-02-000& 01/10/99 02:44 & 28736               & 868                 & 4     & part     \\
30064-01-02-00 & 01/10/99 10:43 & 8947                & 1067                & 4     & no       \\
30064-01-03-00 & 20/11/99 12:21 & 14400               & 594                 & 4     & complete \\
\hline
\end{tabular}
\begin{list}{}{}
\item[$^{\mathrm{a}}$] Total on-source observing time.
\item[$^{\mathrm{b}}$] Average, not corrected for the background, normalised to 5 PCUs.
\end{list}
\end{table*}

\section{Analysis and results}

\subsection{Colour-colour and hardness-intensity diagrams}
We produced
background-subtracted lightcurves binned at 128 s using PCA 
Standard 2 mode data (16 s time resolution). The soft colour (SC) was defined as the ratio of the count rates in the 
bands $3.5-6.4$ keV and $2.0-3.5$ keV, the hard colour (HC) as the ratio of the count rates in the 
bands $9.7-16.0$ keV and $6.4-9.7$ keV. For the intensity, we took the total count rate in the $2.0-16.0$ keV range. Hardness and intensity were
normalised to the Crab values, as obtained during calibration observations close in time and in the same gain epoch \citep[see for details e.g.][]{vanstraaten:2003}.

\begin{figure}
  \resizebox{\hsize}{!}{\includegraphics{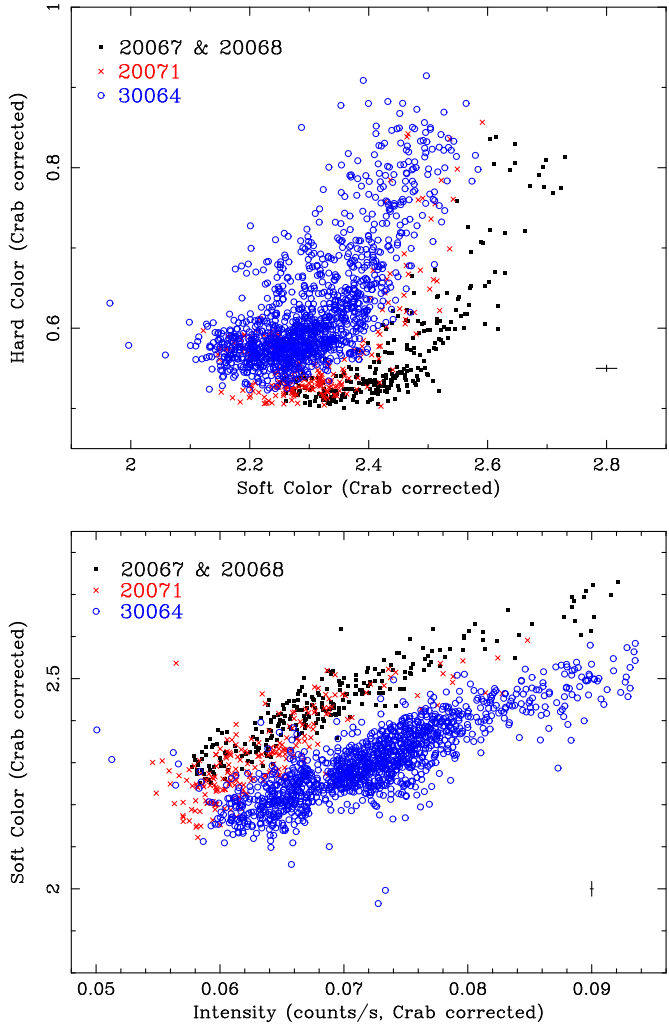}}
  \caption{Colour-colour diagram (upper panel) and hardness-intensity diagram (lower panel) of 4U~1624--49. The soft and hard colours are
  	   defined as the ratio of the count rate in the bands 3.5--6.4 keV/2.0--3.5 keV and 9.7--16.0 keV/6.4--9.7 keV, respectively. The
	   intensity is defined as the source count rate in the energy range 2.0--16.0 keV. The soft and hard
	   colours, as well as the intensity, are normalised to the Crab values. As indicated in each frame, different symbols indicate datasets from
	   different observations, where P20067 and P20068 are taken together. Each data point corresponds to 128 s of data. Typical errors are shown.}
  \label{fig:onverschovenCDplusSID}
\end{figure}

To allow for a better comparison of 4U~1624--49 with known atoll sources, data obtained during dips \citep[see, e.g., Fig.~2 in][]{smale:2001}, where colours and intensity are affected by the obscuration process producing the dips, were excluded from 
the analysis. The resulting CD and HID are shown
in Fig.~\ref{fig:onverschovenCDplusSID}.

We grouped the data into 3 large segments well separated in time.
The observations of P20067 and 
P20068 spanned only 2 days and were taken together in our analysis. 
The result was then compared with the observations of P20071 (spanning 8 days) and 
P30064 (spanning 4 days; 30064-01-03-00 was not used, since it mainly consists of a dip) in the CDs and HIDs.

The shape of the CD suggests that the source is an atoll source in the banana state
\citep{hasinger:1989}. Most of the time, the source is in
the lower banana (softer state); occasionally it
moves through the banana to the harder upper banana region. By investigating the relation between the CD/HIDs and light curves, we found that the softer points in the lower banana are
due to the persistent radiation that the source is normally emitting outside the dips, whereas the harder points come from occasional intensity
flares.

\begin{figure}
  \resizebox{\hsize}{!}{\includegraphics{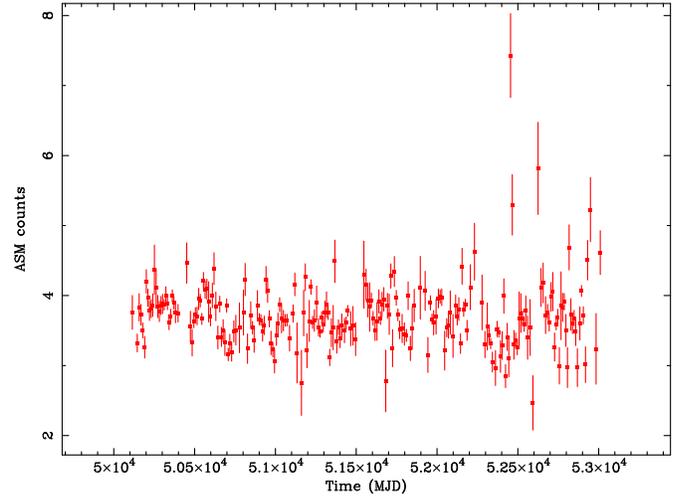}}
  \caption{\emph{RXTE All-Sky Monitor} (\emph{ASM}) lightcurve of 4U~1624--49, showing the day-by-day averages of the \emph{ASM} counts.}
  \label{fig:asm}
\end{figure}

Both the CD and HID show secular shifts \citep[e.g.][]{disalvo:2003} consistent with the SC decreasing, and the HC and/or the intensity gradually increasing over the almost 3-yr period of the observations. The \emph{RXTE All-Sky Monitor} (\emph{ASM}, see Fig.~\ref{fig:asm}) lightcurve shows no clear increasing trend, so we attribute the shift in the HID to a shift in SC. The colours change up to $\sim 10\%$.

\subsection{Power spectra}
\label{sec:power_spectra}
\begin{figure}
  \resizebox{\hsize}{!}{\includegraphics{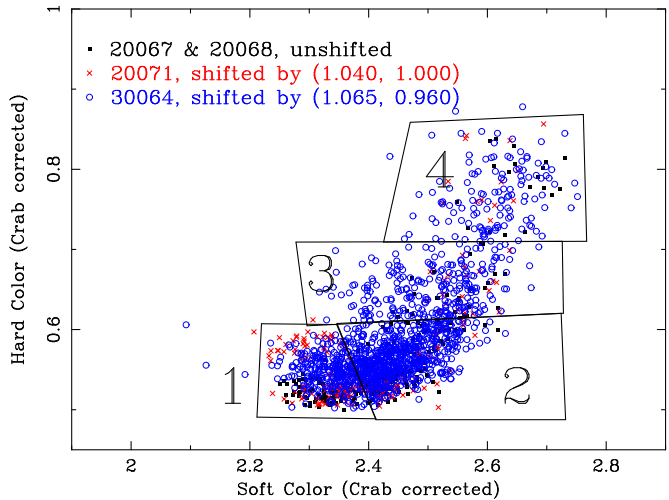}}
  \caption{Same as the upper panel of Fig.~\ref{fig:onverschovenCDplusSID}, but here the data of P20071 and P30064 are shifted so that the tracks corresponding to the different proposals coincide. The selections 1, 2, 3, and 4 are used to create the power spectra.}
  \label{fig:verschovenCDplusSID}
\end{figure}

To study the power spectrum as a function of the position in the CD we created power spectra from all non-dip data obtained in the Good-Xenon mode. Previous experience
with atoll sources shows \citep[e.g.][4U~1636--53]{disalvo:2003} that the power spectra depend on the position of the source relative to the track but not on the track's
location in the CD, i.e., its secular shifts. On this basis we shifted the colour-colour points of P20071 and P30064 to approximately coincide with those of P20067 and P20068.
The data points of P20071 were shifted in SC by 4\%, but not shifted in HC. The data points of P30064 were shifted in SC by 6.5\%, and by -4\% in HC. The result is shown in Fig.~\ref{fig:verschovenCDplusSID}.
We divided the CD into four regions, 1, 2, 3, and 4 (Fig.~\ref{fig:verschovenCDplusSID}). These regions were chosen because the power spectra of atoll sources are observed
to change when the source moves from the lower banana to the upper banana and back \citep[see][]{hasinger:1989}. For statistical reasons, the regions are quite large. A region corresponds to between 142 and 791 power spectra. After
confirming that the power spectra in each region do not depend on the data segment we calculated one average power spectrum for each region.

We used Good Xenon data, with data segments of 128 s and a time
resolution of 1/8192 s ($\sim 125 \mu$s), to produce Fourier power spectra, such that
the lowest available frequency is 1/128 s and the Nyquist frequency is 4096 Hz. 
A Poisson noise level, estimated with the Zhang model
\citep[][]{zhang:1995}, was subtracted. For 4U 1630--47 it was found
\citep[][]{klein-wolt:2004} that the Zhang model does produce the correct
shape for the Poisson noise, but that the estimated level is slightly too
high or too low. To correct for this, an aditional shift was introduced,
where the scaling factor for the shift is determined in a (high) frequency
range where no features are known to be present \citep[see][for the
details of this method]{klein-wolt:2004}. In our analysis we determined
the shift in the 3072--4096-Hz range and found it to be of order $10^{-7}$
in power in all cases, having a negligible effect on the results. We
checked for features in the 3 to 4 kHz range by also creating power
spectra with a Nyquist frequency of 8192 Hz and found none, justifying
this range for the noise estimate.

We then made a fit to the average power spectrum for each region.
We used a power law, in most fits combined with a Lorentzian, to fit the noise below 1 Hz, the ``total VLFN.'' The power-law component is called VLFN as in \citet[][]{hasinger:1989}, and the Lorentzian component is called ``high VLFN'' \citep[see, e.g.,][]{schnerr:2003}. In some cases another significant ($> 3 \sigma$) Lorentzian was detected; this component is called the ``band-limited noise (BLN)'' \citep[see, e.g.,][]{vanstraaten:2002}.
All high-VLFN and BLN features in
the spectra are broad, and could be well described with zero-centered Lorentzians. Therefore we fixed the quality factor $Q$ to $0$. This
results in the characteristic frequency $\nu_{max}$ of each Lorentzian component
being equal to its half-width at half-maximum. The actual Lorentzian functions fitted were of the form
\begin{equation} \label{Lorentzian}
P\left(\nu; \nu_{max}, r\right) = \frac{r^2
\nu_{max}}{\left(\pi/2\right)[\nu_{max}^2 + \nu^2]},
\end{equation}
where $r$ [the fractional root-mean-square (rms) integrated from $0$ to $\infty$], and $\nu_{max}$ were
the independent fit parameters.

\begin{figure*}
  \resizebox{\hsize}{!}{\includegraphics{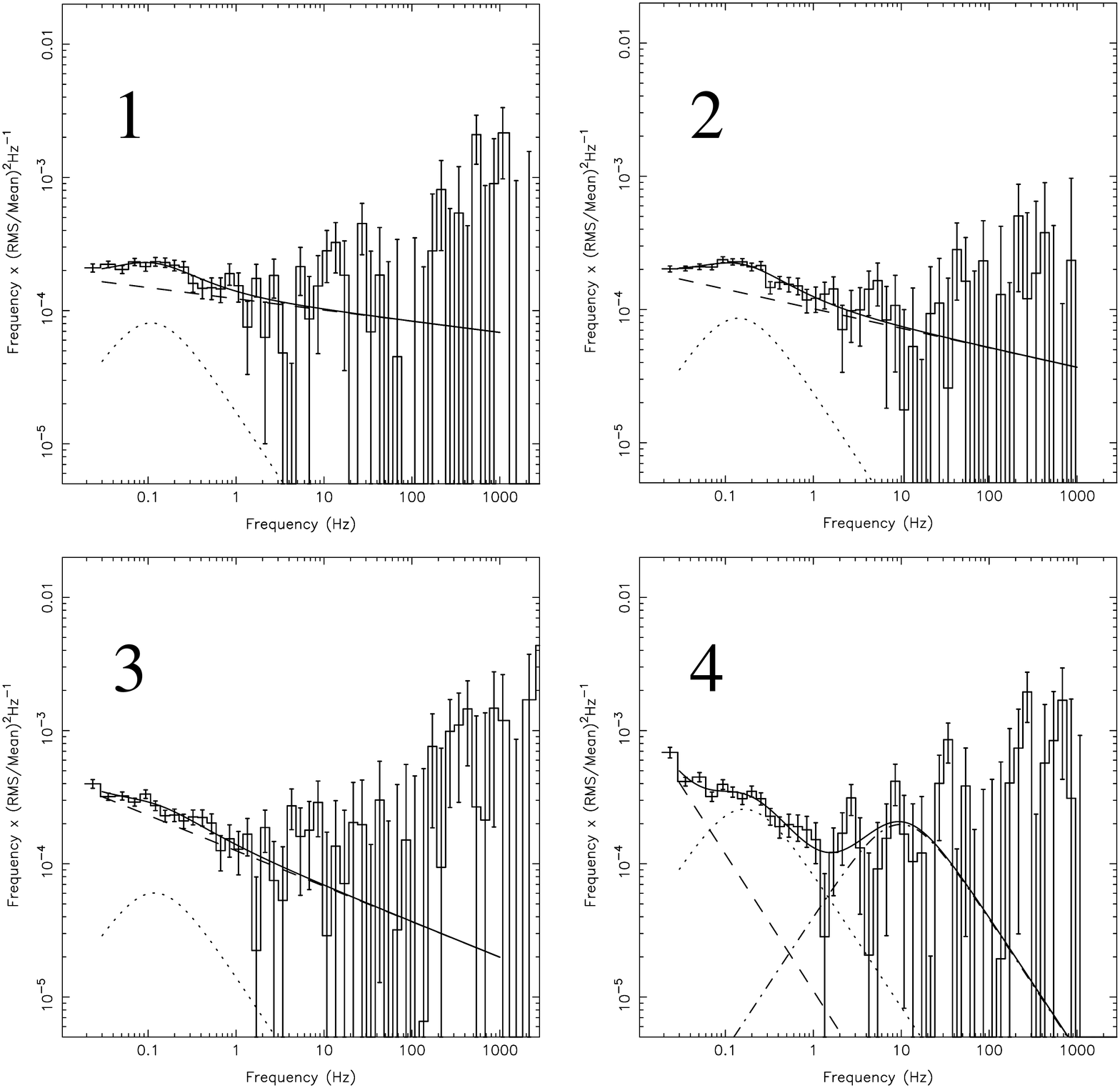}}
  \caption{Power spectra and fit functions in the power-spectral density times frequency representation (see \S~\ref{sec:power_spectra})
  	   of regions 1 to 4 (Fig.~\ref{fig:verschovenCDplusSID}) for fits with a power law plus Lorentzians with only $Q$ of the Lorentzians fixed (at 0). The different components of the
	   fits, where applicable, are as follows: \emph{dashed line}: power law; \emph{dotted line}: first Lorentzian; \emph{dot-dashed line}: second Lorentzian. The sum 
	   of these components, i.e., the full fit, is plotted with a solid line. For the details of the different components, see Table~\ref{tab:fits_table}.}
  \label{fig:fits1234_figure}
\end{figure*}

\begin{table*}
\caption[]{Parameters of the fit components. In all cases, the quality factor $Q$ of the Lorentzians was fixed to $0$. All the values shown are of components detected at better than $3 \sigma$ (based on the error in the power integrated from $0$ to $\infty$), unless otherwise stated.}
\begin{center}
\footnotesize
\label{tab:fits_table}
\vskip 0.2cm
\begin{tabular}{lllll}
\hline \hline
Region					& 1			& 2			& 3				& 4			\\
\hline
Rate$^{\mathrm{a}}$ ($10^2$ c/s)	& $7.1 \pm 0.3$		& $7.1 \pm 0.3$		& $6.9 \pm 0.3$			& $8.2 \pm 0.3$		\\
$\#$ power spectra			& 416			& 791			& 262				& 142			\\
Background$^{\mathrm{b}}$ ($10^2$ c/s)	& $1.17 \pm 0.11$	& $1.03 \pm 0.10$	& $0.93 \pm 0.09$		& $1.00 \pm 0.10$	\\
\hline
\multicolumn{5}{c}{all parameters left free}													\\
\hline
VLFN $\alpha$				& $1.09 \pm 0.03$	& $1.14 \pm 0.03$	& $1.27 \pm 0.03$		& $2.0 \pm 0.2$		\\
VLFN rms$^{\mathrm{c}}$ (\%)		& $2.62 \pm 0.10$	& $2.59 \pm 0.07$	& $3.35 \pm 0.05$		& $3.44 \pm 0.15$	\\
high VLFN $\nu_{max}$			& $0.11 \pm 0.02$	& $0.14 \pm 0.02$	& $0.12$			& $0.16 \pm 0.03$	\\
high VLFN rms (\%)			& $1.58 \pm 0.16$	& $1.61 \pm 0.12$	& $< 1.9^{\mathrm{d}}$		& $2.84 \pm 0.22$	\\
total VLFN rms (\%)			& $3.06 \pm 0.06$	& $3.05 \pm 0.05$	& $3.3 < rms < 5.3$		& $4.47 \pm 0.06$	\\
BLN $\nu_{max}$				& -			& -			& -				& $10 \pm 6$		\\
BLN rms (\%)				& -			& -			& -				& $2.5 \pm 0.4$		\\
$\chi^2/d.o.f.$				& 120/95		& 79/95			& 102/97			& 104/93		\\
\hline
\multicolumn{5}{c}{power law only}														\\
\hline
$\alpha$				& $1.065 \pm 0.016$	& $1.084 \pm 0.012$	& $1.22 \pm 0.02$		& $1.32 \pm 0.03$	\\
rms$^{\mathrm{c}}$ (\%)			& $3.02 \pm 0.03$	& $3.00 \pm 0.02$	& $3.61 \pm 0.04$		& $4.18 \pm 0.06$	\\
$\chi^2/d.o.f.$				& 145/95		& 129/95		& 100/95			& 127/95		\\
\hline
\multicolumn{5}{c}{fixed power law}														\\
\hline
VLFN $\alpha$				& 1.7 (fixed)		& 1.7 (fixed)		& 1.7 (fixed)			& 1.7 (fixed)		\\
VLFN rms$^{\mathrm{c}}$ (\%)		& $1.68 \pm 0.09$	& $1.82 \pm 0.06$	& $2.76 \pm 0.10$		& $3.65 \pm 0.11$	\\
high VLFN $\nu_{max}$ (Hz)		& $0.107 \pm 0.015$	& $0.131 \pm 0.011$	& $0.14 \pm 0.02$		& $0.20 \pm 0.03$	\\
high VLFN rms (\%)			& $2.39 \pm 0.08$	& $2.38 \pm 0.06$	& $2.39 \pm 0.09$		& $2.37 \pm 0.12$	\\
total VLFN rms (\%)			& $2.93 \pm 0.04$	& $2.99 \pm 0.03$	& $3.65 \pm 0.04$		& $4.35 \pm 0.04$	\\
BLN $\nu_{max}$ (Hz)			& $1.6 \pm 0.8$		& $2.6 \pm 1.0$		& $6 \pm 3$			& $12^{+12}_{-5}$	\\
BLN rms (\%)				& $1.89 \pm 0.16$	& $1.79 \pm 0.15$	& $2.3 \pm 0.3$			& $2.5 \pm 0.5$		\\
$\chi^2/d.o.f.$				& 130/94		& 81/94			& 104/94			& 107/94		\\
\end{tabular}
\end{center}
Errors correspond to $\Delta \chi^2 = 1$.
\begin{list}{}{}
\item[$^{\mathrm{a}}$] PCA count rate, 5 PCUs, not corrected for the background. Counting statistics errors are quoted.
\item[$^{\mathrm{b}}$] Averaged background. Counting statistics errors are quoted.
\item[$^{\mathrm{c}}$] rms amplitude calculated in the frequency range 0.01--1.0 Hz.
\item[$^{\mathrm{d}}$] 95\% confidence upper limit.
\end{list}
\end{table*}

We plot the power spectra and the fit functions in the power times frequency
representation where power
spectral density is multiplied with its Fourier frequency. 
 The results are shown in
Fig.~\ref{fig:fits1234_figure}
and in Table~\ref{tab:fits_table}. All power spectra are dominated by the VLFN, which is fitted with a power law and a Lorentzian. The Lorentzian in region 3 is not significant. For this feature we calculated an upper limit to the rms. Also the total VLFN rms was calculated and is shown in Table~\ref{tab:fits_table}. The feature in region 3 around 3000 Hz is not real; it is due to aliasing and disappears when a Nyquist frequency of 8 kHz instead of our usual 4 kHz is used.
In region 4,
we found a significant feature around 10 Hz which we attribute to the 
BLN. There seems to be a feature around 3 Hz as well, but this feature is not significant.

The steepness of the power law fitting the VLFN increases when the source moves from region 1 through to 4. This is what we expect from an atoll source in the banana state.
Also the fractional rms is consistent with increasing as the hardness increases. The Lorentzian at 10 Hz in region 4 affects the index and rms of the power law only slightly. Fitting this region with a power law and only one Lorentzian gives %$\alpha = 1.32 \pm 0.03$ and an rms of $4.2 \pm 0.7\%$
$\alpha = 1.8 \pm 0.3$ and an rms of $1.8 \pm 0.3\%$, making this still the steepest and strongest power law of the four. We also made fits to the power spectra using a power law and no Lorentzians. This allows for a better comparison between region 3, which has no significant Lorentzians, and the other regions. The results 
of this are shown 
in 
Table~\ref{tab:fits_table}. Also in this case steepness and rms increase as the source moves 
from 1 to 4. The fact that $\alpha$ does not increase when moving from region 1 to 
2, may be a result of the choice of the regions, i.e., region 2 has the same hard 
colour as region 1.
Finally we fitted the power spectra with the power-law steepness fixed to
$\alpha = 1.7$, thus assuring that the power law fitting the VLFN does not fit
to features at higher frequencies. 
We then find two Lorentzians in each region. The results are shown in Table~\ref{tab:fits_table}. The rms of the power law still increases when the source moves to the harder part of the CD. Besides we note that the characteristic frequencies of the Lorentzians significantly increase when the source moves from region 1 to region 4, from $0.11 \pm 0.02$ to $0.20 \pm 0.03$ and from $1.6 \pm 0.8$ to $12^{+12}_{-5}$ for the high VLFN and the BLN, respectively.

\section{Discussion}
\label{sec:discussion}
% DISCUSSION
We analysed $\sim 360$ ks of archival \emph{RXTE} data of 4U~1624--49 taken between January 1997
and November 1999.

The shape of the CD suggests that 4U~1624--49 is an atoll source, which
is in the banana state at the time of the observations.
We will investigate this suggestion in the light of the paper by \citet[][]{hasinger:1989} in which the two classes of LMXBs, atoll and Z sources, are first defined. They find that atoll sources differ from Z sources in:
\begin{itemize}
  \item the absence of horizontal branch QPOs,
  \item the fact that the VLFN is less steep,
  \item the presence of clear wiggles in the VLFN,
  \item the fact that they often show very strong high-frequency noise \citep[BLN in this paper, after e.g.][]{vanstraaten:2002}, which is arguably different from normal-branch/flaring-branch QPOs in Z sources.
\end{itemize}
We first note that no QPO is detected in the power spectra of 4U~1624--49 \citep[see also][]{smale:2001}, and that the power spectra also show wiggles (the high-VLFN components) in the VLFN. \citet[][]{balucinska:2001} argued that during flare evolution systematic changes clearly take place in the HIDs very similar to the flaring branch of Z-track sources displayed on similar diagrams.
\citeauthor[][]{hasinger:1989} find a VLFN power-law index in the range $1.3 < \alpha < 1.9$ on the flaring branch, and $0.9 < \alpha < 1.5$ on the banana branch. When we 
leave the power-law index free in our fits, we find it to be in the range $1.0 < \alpha < 1.4$ (except in the case where also a significant BLN feature is found, see 
Section~\ref{sec:power_spectra}). From these arguments we conclude that 4U~1624--49 should be classified as an atoll source in the banana branch. 
Consistent with this, the values of $\alpha$ and the fractional rms of the VLFN tend to increase when moving to the upper right in the CD, like other atoll sources 
moving up the banana branch.

\begin{figure*}
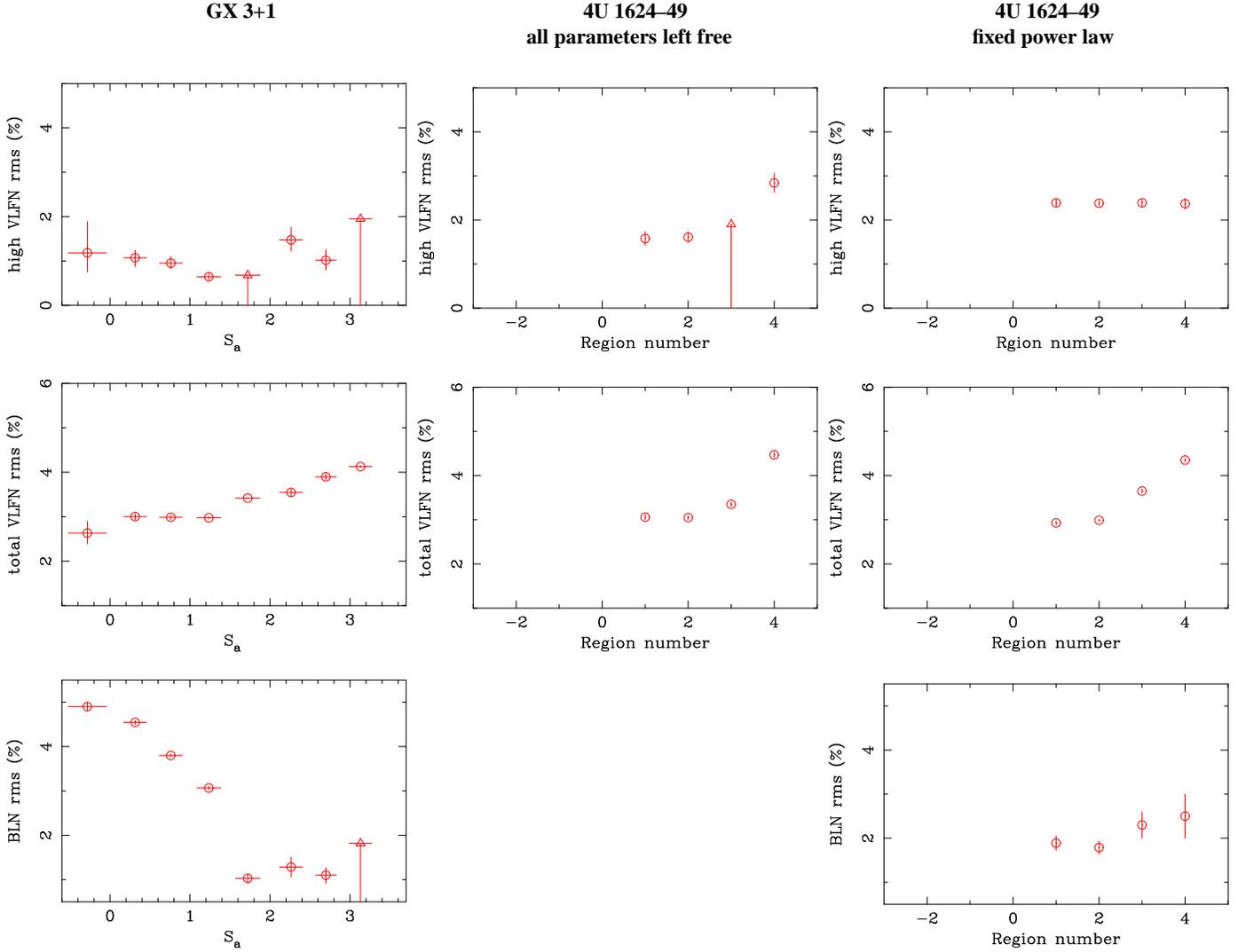

\hfill\parbox[b]{15cm}{\bf{GX~3+1}\hspace{4.6cm}\bf{4U~1624--49}\hspace{4.6cm}\bf{4U~1624--49}}
\hfill\parbox[b]{18cm}{\hspace{7.8cm}\bf{all parameters left free}\hspace{3.6cm}\bf{fixed power law}}
\begin{flushleft}
\vspace{-0.1cm}
\hspace{0cm}
\hbox{\psfig{figure=0405fig5.ps,width=6cm,angle=270}
\hspace{0cm}
\psfig{figure=0405fig6.ps,width=6cm,angle=270}
\hspace{0cm}
\psfig{figure=0405fig7.ps,width=6cm,angle=270}}
\vspace{0cm}
\hspace{0cm}
\hbox{\psfig{figure=0405fig8.ps,width=6cm,angle=270}
\hspace{0cm}
\psfig{figure=0405fig9.ps,width=6cm,angle=270}
\hspace{0cm}
\psfig{figure=0405fi10.ps,width=6cm,angle=270}}
\vspace{0cm}
\hspace{0cm}
\hbox{\psfig{figure=0405fi11.ps,width=6cm,angle=270}
\hspace{0cm}
\hspace{6cm}
\hspace{0cm}
\psfig{figure=0405fi12.ps,width=6cm,angle=270}}
\hfill\parbox[b]{18cm}{\caption[]{Comparison between the trends in the high VLFN rms, the total VLFN rms, and the BLN rms of GX~3+1 and 4U~1624--49. The values of 
GX~3+1 are taken from \citet{reerink:2005} and are representative for the bright GX sources GX~9+9, GX~9+1, and GX~3+1. Note that \citet[][]{reerink:2005} use so-called 
$S_a$ values to define their intervals, where $S_a > 1.5$ corresponds to GX~3+1 being in the UB. We plotted our results on a similar scaling for easier comparison. 
Because no significant BLN component is found for 4U~1624--49 in three of the four regions when all parameters are left free in the fit, we also show the values for the fits 
in which the power-law index was fixed. We find that based on these criteria 4U~1624--49 is in the UB throughout the observations.}\label{fig:comparison_with_reerink}}
\end{flushleft}
\end{figure*}

The banana branch is usually subdivided into the lower left banana
(LLB), the lower banana (LB), and the upper banana \citep[UB, see again][for definitions]{hasinger:1989}. 
\citet{reerink:2005} in their study of GX~3+1, GX~9+1, and GX~9+1 propose to define as a boundary between the LB and the UB the point where the trends of the power spectral 
components as a function of position in the branch change. The most notable trend changes in their study are those in the high-VLFN rms, which changes from a flat or slightly 
decreasing trend to an increasing trend, and that in the BLN rms, which changes from a decreasing trend to a flat or slightly increasing trend. Comparing our results to 
theirs (see Fig.~\ref{fig:comparison_with_reerink}), we find that 4U~1624--49 is in the UB.

In regions 1, 2, and 4, we found a significant (larger than 3$\sigma$) Lorentzian with a characteristic frequency in the range of $\sim$0.11 to $\sim$0.16 Hz, a high-VLFN component. The fractional rms of this Lorentzian ranged from $\sim$1.6\% to 2.8\%. For region 3, we could set an upper limit to a similar Lorentzian with a fractional rms of 1.9\% (95\% confidence). These bumps in the VLFN are probably the same as those reported for the bright atoll sources GX 13+1 \citep[][$\sim$3\% rms at 1 Hz]{schnerr:2003}, and e.g. GX~9+9 and GX~9+1 \citep[][$\sim$2\% up to $\sim$4\% at $\sim$0.1 Hz]{reerink:2005}.
We also found a significant Lorentzian at 10 Hz in region 4.
This is curious, since for other atoll sources, the BLN component at higher frequencies becomes weaker when the source moves to
the upper banana. On the other hand, some atoll sources in the banana state show a QPO around 6 Hz when the source is at
the tip of the upper banana. We searched for this QPO by applying the analysis outlined in \S~\ref{sec:power_spectra} to the top-most part of region 4, but did not find it.

When using a power law with index $\alpha$ fixed to $1.7$ to fit the ``low'' VLFN, all four regions show a significant BLN component. However, whereas for other sources the characteristic frequency of the BLN tends to decrease as the source moves to the upper banana, in 4U~1624--49 this frequency increases. Also, the BLN frequency never reaches values of several 10 Hz, such as in most of the atoll sources
(including GX 13+1 \citep[][up to 30 Hz]{schnerr:2003}, and GX 9+9 and GX 3+1 \citep[][up to 60 Hz and 24 Hz, respectively]{reerink:2005}).

We observed secular shifts in the colour-colour and hardness-intensity diagrams.
Shifts like these are also observed in some Z sources \citep[][ and references therein]{smale:2003,kuulkers:1994,kuulkers:1996}, and in the atoll
sources 4U 1636--53 \citep{prins:1997,disalvo:2003} and GX 13+1 \citep{schnerr:2003}. 
The shifts in 4U~1624--49 are most clear in the colours, up to $\sim 10\%$, with little difference between HC and SC. The intensity of 4U~1624--49 is rather stable, as seen with the \emph{ASM} (Fig.~\ref{fig:asm}). In 4U 1636--53, the shifts in colours are only a few per cent, whereas the shift in intensity is up to $\sim 20\%$. The shifts in GX 13+1 are up to $\sim 40\%$ in HC, $\sim 7\%$ in SC, whereas there is only a shift in intensity in the upper-most part of the banana (up to $\sim 20\%$). The shifts in 4U~1624--49 towards higher HC and/or lower SC seem to be rather gradual over the almost 3-yr period of the observations, whereas GX 13+1 moves back and forth between higher and lower HC on a time scale of weeks. The shift in 4U 1636--53 is not gradual at all, this source appears to ``jump'' from one state to another as is reflected in the lower kHz QPO frequency vs. count rate diagram \citep[Fig. 2 in][]{disalvo:2003}. This may indicate a different source for the secular shifts in 4U 1636--53 as opposed to GX 13+1 and 4U~1624--49. \citet{schnerr:2003} propose secular geometry changes such as precession as possible sources for the secular shifts, which would account for the gradual change. They assume a head-on view to GX 13+1 (and a jet scenario) to explain the large shift in HC. 4U~1624--49 is a dipping source and thus has a high inclination \citep[$60^\circ \lesssim i \lesssim 75^\circ$ according to][]{frank:1987}, which might explain the much smaller shift in HC for this source. For more insight in the processes behind the secular shifts, more data on different sources and per source is necessary, and thus long-term X-ray observations of several sources are recommended.

\section{Conclusion}
\label{sec:conclusion}
Outside the dips, 4U~1624--49 shows properties that are mostly consistent with those of an atoll source at high luminosity: a ``middle'' and upper banana branch in CD/HIDs, secular shifts of the tracks, a power spectrum dominated by a few $\%$ rms VLFN that becomes stronger and steeper at harder colours. All this is very similar to what is observed in ``GX atoll sources'' such as GX~3+1 and GX~9+9 \citep{reerink:2005}. However, to the extent that it can be detected the characteristic frequency of the BLN may increase, rather than decrease as in those sources. During our observations, the luminosity of 4U~1624--49 was $(1.5-1.7) \times 10^{38}$ ergs s$^{-1}$ \citep{smale:2001,balucinska:2001}, i.e., $\sim 0.5-0.8 L_{\rm Edd}$, which is in accordance with luminosities inferred for the GX atoll sources. Combining all this with the fact that 4U~1624--49 shows periodic dips and is rather stable in flux we conclude that 4U~1624--49 most likely is a GX atoll source seen edge on. The fact that the BLN in this source has a characteristic frequency that is atypically low and monotonically increasing may be related to this unusual inclination of the source, perhaps because the innermost region of the disk where the highest frequencies originate is obscured by structure slightly further out, e.g. a puffed-up disk region associated with the radiation-pressure dominated region.%This interpretation may also be an indication for the inclination of GX 9+1 to be somewhat higher than that of GX 3+1, GX 9+9 and GX 13+1, since the BLN characteristic frequency of GX 9+1 also stays below 13 Hz \citep{reerink:2005}.

\begin{acknowledgements}
  We would like to thank our colleagues of The X-ray/high-energy astrophysics group at the ``Anton Pannekoek Institute'' for their help and useful discussions. DL especially wishes to thank Thomas Reerink. This work was supported in part by the Netherlands Organisation for Scientific Research (NWO). This research has made use of data obtained through the High Energy Astrophysics
  Science Archive Research Center Online Service, provided by the NASA/Goddard Space Flight Center.
\end{acknowledgements}

\bibliographystyle{aa}
\bibliography{references}

\end{document}